\documentstyle[aas2pp4]{article}
\received{}
\revised{}
\accepted{}
\righthead{UV and Optical Galaxy Morphology}
\lefthead{Kuchinski et al.}

\begin{document}

\title{Comparing Galaxy Morphology at Ultraviolet and Optical Wavelengths}

\author{L. E. Kuchinski\altaffilmark{1}, W. L. Freedman\altaffilmark{2}, Barry F. Madore\altaffilmark{1,2}, M. Trewhella\altaffilmark{1}, R. C. Bohlin\altaffilmark{3}, R. H. Cornett\altaffilmark{4}, M. N. Fanelli\altaffilmark{4,5}, P. M. Marcum\altaffilmark{6}, S. G. Neff\altaffilmark{7}, R. W. O'Connell\altaffilmark{8}, M. S. Roberts\altaffilmark{9}, A. M. Smith\altaffilmark{7}, T. P. Stecher\altaffilmark{7}, W. H. Waller\altaffilmark{4,10}}
\altaffiltext{1}{Infrared Processing and Analysis Center, Caltech/JPL, Pasadena, CA 91125}
\altaffiltext{2}{Observatories of the Carnegie Institution of Washington, Pasadena, CA 91101}
\altaffiltext{3}{Space Telescope Science Institute, Baltimore, MD 21218}
\altaffiltext{4}{Raytheon ITSS Corp., NASA Goddard Space Flight Center, Greenbelt, MD 20771}
\altaffiltext{5}{Department of Physics, University of North Texas, Denton, TX 76023}
\altaffiltext{6}{Department of Physics, Texas Christian University, Fort Worth, TX 76129}
\altaffiltext{7}{Laboratory for Astronomy and Solar Physics, NASA Goddard Space Flight Center, Greenbelt, MD 20771}
\altaffiltext{8}{Department of Astronomy, University of Virginia, Charlottesville, VA 22903}
\altaffiltext{9}{National Radio Astronomy Observatory, Charlottesville, VA 22903}
\altaffiltext{10}{Department of Physics and Astronomy, Tufts University, Medford, MA 02155}

\begin{abstract}
We have undertaken an imaging survey of 34 nearby galaxies in
far--ultraviolet (FUV, $\sim 1500 \AA$)
and optical ($UBVRI$) passbands to characterize galaxy morphology as a function
of wavelength.
This sample, which includes a range of classical Hubble types from 
elliptical to irregular with emphasis on spirals at low inclination angle,
provides  a valuable database for comparison 
with images of high--$z$ galaxies whose FUV light is redshifted into the optical
and near--infrared bands.
Ultraviolet data are from the UIT {\it Astro--2} mission.
We present images and surface brightness profiles for each galaxy, and we
discuss the wavelength--dependence of morphology for different Hubble types
in the context of understanding high--$z$ objects.
In general, the dominance
of young stars in the FUV produces the patchy appearance
of a morphological type later than that inferred from optical images.
Prominent rings and circumnuclear star formation
regions are clearly evident in FUV images of spirals,
while bulges, bars, and old, red stellar disks
are faint to invisible at these short wavelengths.
However, the magnitude of the change in apparent morphology ranges from 
dramatic in early--type spirals with prominent optical bulges
to slight in late--type spirals and irregulars, in which
young stars dominate both the UV and optical emission.  
Starburst galaxies with centrally concentrated, symmetric bursts display
an apparent ``E/S0'' structure in the FUV, 
while starbursts associated with rings or mergers produce a peculiar 
morphology.
We briefly discuss the inadequacy of the optically--defined Hubble sequence to 
describe FUV galaxy images and estimate morphological $k$--corrections, and we
suggest some directions for future research with this dataset.
\end{abstract}

\section{Introduction}
 
\par
An understanding of the ultraviolet (UV) properties of local galaxies is
essential for interpreting images of high--redshift systems and especially 
critical in searches for
morphological evolution in the galaxy population.
At redshifts of $ z \sim 3 - 10$, the rest--frame UV light of 
galaxies is shifted into the optical (1500$\AA$ observed in $V$--band
at $z \sim 2.7$) and near--infrared (1500$\AA$ observed in $H$--band at $z \sim 10$) regions of the 
spectrum that are easily observable from the ground.  Although one operational
approach to overcoming this ``band--shifting'' problem is to observe
in the infrared
({\it e.g.} \cite{tep98}), 
the rate of data acquisition for galaxy images is generally highest at optical
wavelengths where detectors are larger and the sky background is lower 
than in the infrared.  Under these circumstances, the most direct method 
to compare
present--day galaxies to those in the early universe is to build a large
sample of local UV galaxy data.

\par
Recent deep surveys suggest that galaxies with peculiar morphology are 
more prevalent at high redshift ({\it e.g.} \cite{bri98}; \cite{abr96a}).
However, it is difficult to quantify the
influence of band--shifting on observed differences between the optical
characteristics of nearby galaxies and the rest--frame UV appearance of
distant ones because UV and optical morphology are not often well--coupled
(\cite{oco97a}; Marcum et al. 1997, 2000).  Restframe UV images of galaxies
often suggest a later Hubble type than the
corresponding optical appearance:  bulges are less prominent, galaxy light
appears more patchy, and in extreme cases the different star--forming regions
in a single galaxy may appear as separate systems in the UV (\cite{oco97a}; 
\cite{oco96}). On the other hand, recent deep NICMOS imaging of distant
galaxies at longer rest wavelengths suggests that some of the morphological
peculiarities are still visible in the optical and thus are intrinsic in
the galaxies' structure ({\it e.g.} \cite{bun99}; \cite{tep98}).
Although methods exist to estimate UV 
morphology by extrapolation from optical data (\cite{abr98}; \cite{afm97}),
they rely on
the use of global galaxy spectral energy distributions  that may not accurately
represent localized conditions, especially in dusty or starbursting regions.  
For example, Donas, Milliard, \& Laget (1995) find a variation among UV$-b$
colors in galaxies with identical $b-r$.
The availability of FUV images for bright, well--resolved galaxies of a 
range of Hubble types would facilitate a more direct comparison between local
and distant galaxy populations.

\par
The appearance of galaxies in the UV is determined primarily by emission from
hot stars and by the distribution of dust, with a contribution in the central
regions from an AGN if one is present and unobscured.
Light in the $\sim$1500$\AA$ far--ultraviolet (FUV) band of the 
Ultraviolet Imaging Telescope (UIT, \cite{ste97})
originates in young objects of spectral types O and B (Fanelli et al. 1997)
and, if old populations are present, in certain types of low--mass stars 
in late stages of evolution (the ``UVX'' population, see \cite{oco99} for a review). 
Thus this region of the spectrum can probe timescales 
of young stellar populations between that studied via $H\alpha$ emission
(only O stars, $\sim$ 5 Myr) and that evident in the optical colors (which
trace the age of a stellar population on timescales of a
few Gyr; O'Connell 1997b) .

\par
Opacity due to dust in the FUV is a factor of $\sim$2.5 higher
than in the optical $V$ band, assuming a Galactic extinction law 
(\cite{ccm89}), and even larger for Small Magellanic Cloud--type dust
 (\cite{gor97}).  The young, massive stars that provide the bulk of the
UV emission also tend to have small scale heights, less than or similar to that
of the dust, while older stars that dominate at optical wavelengths and 
have scale heights greater than that of the dust layer (\cite{mih81}).
The degree to which dust extinguishes the UV light of
highly inclined disk galaxies is unclear. It likely depends strongly on
the relative
geometry of star--forming regions and dust concentrations, a phenomenon
recently noted in optical/NIR studies of nearby galaxies as well as in
theoretical studies of systems with a mixture of stars and dust
(\cite{kuc98}; Gordon et al. 1997; \cite{hui94}).  For instance,
NGC 4631 is quite bright in the FUV and blue in FUV--optical colors, and it 
does not show evidence for strong
attenuation (Smith et al. 1997, 2000; \cite{oco97a}). 
However, NGC 891 is only barely detected 
(Marcum et al 1997, 2000), a phenomenon that Smith et al. suggest is due 
to extinction effects.  UV morphology may also be dramatically influenced by 
specific features such as dust lanes, as noted in Cen~A 
by O'Connell (1997a).    
Scattering by dust grains is quite efficient in the FUV (\cite{wit92})
and may contribute significantly to the emission near H II regions and
in spiral arms (\cite{wal97a}; \cite{ste82}).
\par
Prior to the Space Shuttle--borne UIT {\it Astro} missions,
UV imaging of galaxies was
limited to low resolution and/or small fields of view (see \cite{bro99} for
a detailed review).  UIT--prototype 
rocket--borne experiments
in the 1980's provided images of five nearby galaxies with a  resolution
of $\sim$15$\arcsec$  and ten more with a resolution of 100$\arcsec$
(\cite{hil84}; \cite{boh91}). The FOCA balloon
experiment yielded images of similar resolution for six nearby galaxies
(\cite{ble90}).  Recently, higher resolution UV images of 110 galaxy 
nuclei (\cite{mao96}) and nine starburst galaxies (\cite{meu95}) have been
obtained at 2200$\AA$ using the Hubble Space Telescope's
Faint Object Camera with a very small (22\arcsec) field of view. 
A FUV ($\sim$1500$\AA$) survey of galaxies at redshifts of
$\sim$~0.1--0.2 is also underway
using the Space Telescope Imaging Spectrograph, again with a small (25\arcsec)
field of view (\cite{gar97}). 
With a 40$\arcmin$ field of view and $\sim$3$\arcsec$ 
resolution, UIT was optimally suited to 
study the morphology of nearby galaxies with good resolution.
The usefulness of UV and optical images and radial profiles to study 
morphology variations with wavelength was demonstrated  by \cite{cor94}
 with UIT {\it Astro--1}
data for M74,  by \cite{hil97} for M51, and  by \cite{rei94} for M81.

\par
In this and a companion paper (Marcum et al. 2000), we present
the results of an UV--optical imaging survey of 
nearby bright galaxies.  The UV images were obtained by the UIT during the 
{\it Astro--1} and {\it Astro--2} Spacelab missions in 1990 and 1995 
respectively.  Ground--based optical CCD images of the sample galaxies 
were obtained with pixel scales and fields of view approximately comparable
to those of the UIT images.  The {\it Astro--1} data presented in 
Marcum et al. include both FUV ($\sim$1500$\AA$) and NUV($\sim$2500$\AA$)
data as well as optical images for 43 galaxies.  This paper includes FUV
and optical data for 34 galaxies observed during {\it Astro--2}
with Hubble types specifically chosen to span
the range E to Irr, including some interacting/merging systems.  
There is some emphasis in the sample selection on face--on spirals in 
which spiral arm, bar, and ring morphologies are easily observed.
In Section 2 we discuss the sample selection, describe the UIT
and optical observations, and explain the basic data reduction.
Section 3 contains details of the registration of FUV and optical images and
the extraction of surface brightness and color profiles.  A qualitative
comparison of UV and optical morphology, based on the images and profiles,
is presented in Section 4.  In Section 5 we provide a brief summary and 
discuss avenues of future investigation, including studies of the 
morphological $k$--correction and analysis of the star--formation histories
of individual galaxies.  In a subsequent paper (\cite{kuc00}), we will
investigate the quantitative concentration and asymmetry indices for these
galaxies and consider the implications of these results for the study of 
high--redshift galaxies.

\section{Observations and Data Reduction}

\subsection{Sample Selection}

\par
Our sample consists of 34 galaxies observed with UIT during the
{\it Astro--2} mission for which we have also obtained ground--based optical
images.  Many of the galaxies were selected as part of a UIT Guest Observer
program specifically to investigate the UV morphology in prototypes
of different Hubble types.  Others were observed during UIT studies of 
starburst galaxies ({\it e.g.} \cite{smi96}), early--type galaxies 
(\cite{ohl98}), AGNs (Fanelli et al. 1997a), or individually for
selected purposes.  It is important to remember that while this sample does
contain galaxies with a range of morphologies, it was not chosen to 
statistically represent the relative distribution of Hubble types 
in the local population. 
The galaxies have types E to Irr ($T$ = --5.0 -- +10.0), and are located
at distances of $\sim 2 - 25$ Mpc ($H_0 = 75$km/s/Mpc).
Late--type systems in which star
formation produces prominent morphological features in the FUV are emphasized.
Three sets of interacting pairs are included:  NGC 3226/7, NGC 4038/9, and
NGC 5194/5. Table~\ref{galxdat} gives basic properties of the sample galaxies.

\subsection{Ultraviolet Observations and Data Reduction}

\par
The sample galaxies were observed in the FUV with UIT, a 38cm. 
Ritchey--Chretien telescope mounted on the {\it Astro} payload.  The
{\it Astro--2} mission during which these data were obtained was flown on
the Space Shuttle Endeavour on 2--18 March 1995.
Details of the telescope and instrumentation, as
well as specific information about pipeline data processing of the UV images, 
can be found in Stecher et al. (1997) and will be briefly summarized here.
Only the FUV ($\sim $1500\AA) camera was operational during the 
{\it Astro--2} mission.
Two FUV filters were used:  the B1 filter with an effective wavelength of
1520${\rm \AA}$ and width 354${\rm \AA}$, and the slightly narrower
B5 filter with effective wavelength of 1615${\rm \AA}$ and width
225${\rm \AA}$, which was used during
daylight observations to exclude dayglow emission lines (\cite{hil98};
\cite{wal95}).
Based on observations of early--type galaxies (\cite{ohl98}), 
there is no evidence for a ``red leak'' in the system.  The camera consisted
of a two--stage image intensifier coupled to Kodak IIa--0 film on which
the images were recorded.  The photographic film was digitized 
with a 20$\mu$m square aperture and 10$\mu$m sample spacing, then binned
to a 20$\mu$m spacing (\cite{ste97}).  In the pipeline data reductions,
background ``fog'' from the photographic film was subtracted, and the 
images were linearized, flatfielded, and calibrated.  The final output images
have a scale of 1.136\arcsec/pixel and a typical point spread function of
FWHM $\sim$ 3$\arcsec$ in the central 16$\arcmin$ of the frame, in which nearly
all of the galaxies studied here are imaged (\cite{ste97}).
UIT images with enough stars to calculate astrometric solutions were corrected
for distortion (\cite{ste97}; \cite{gre94}); these
data products were used where available.  However, we note that even on 
uncorrected images, the uncertainty due to distortion
in the central part of the frame (where most of our galaxy images are located)
is less than or equal to the PSF FWHM.

\par
After visual inspection of all available FUV images for each galaxy, we 
selected the ones with the best signal--to--noise and image quality for further 
analysis.  In most cases this is the image with the longest exposure time, 
unless that image is noted in the UIT log to suffer 
pointing problems or other defects. In such cases, the longest
exposure time image of good quality was used. 
The UV--bright center of NGC 4151 is saturated in the UIT {\it Astro--2} 
images; for photometry of this galaxy, see \cite{fan97a}.
We select a long--exposure UIT image of this galaxy to show
the faint spiral arms as well as the active nucleus.
The UIT exposure time of the image selected for each galaxy
is given in Table~\ref{galxd2}.

\par
Many of the FUV images required no further processing beyond
the UIT pipeline, but some have a noticeable ``stripe'' artifact through
the center of the frame (\cite{ste97}).  
If the entire galaxy was located within the stripe, no
correction was made and care was taken to determine the sky value within the
stripe as well.  For galaxies located partly within and partly outside the
stripe, a quadratic surface was fit to the frame after masking out the 
central region (where the galaxy lies) and the frame edges.
This surface was then subtracted before the sky background and surface
brightness profile were measured.  The FUV sky background is typically low
(see \cite{wal95} for a discussion of the UV background),
ranging from undetectable (with a 1--$\sigma$ limit of 
$\sim$25~mag~arcsec$^{-2}$) up to 22~mag~arcsec$^{-2}$ for a few images
acquired during daylight.   

\subsection{Optical Observations and Data Reduction}

\par
Broad--band optical images of the sample galaxies were obtained over
the past several years with the Las Campanas Observatory 2.5m du Pont 
telescope, the CTIO 1.5m telescope, and the Palomar 1.5m and 5m telescopes.  
All of the galaxies presented here were imaged in at least one of the 
$UBVR_{c}I_{c}$ bands; many have multi--color photometry.
Observations at Las Campanas were carried out using the 2048$\times$2048
Tek 5 CCD with a scale of 0.26\arcsec/pixel.  The CTIO observations 
utilized the Tek 2048$\times$2048 CCD with a scale of 0.43\arcsec/pixel.
Data from the Palomar 1.5m telescope were taken using the 2048$\times$2048
CCD13 or CCD16 with a scale of 0.37\arcsec/pixel; some of these were 
binned to 1024$\times$1024 with a scale of 0.74\arcsec/pixel. Observations
with the Palomar 5m telescope used the COSMIC camera with a 2048$\times$2048
CCD and a scale of 0.28\arcsec/pixel; again some images were binned
to a scale of 0.55\arcsec/pixel.
Table~\ref{galxd2} shows the telescope, filters, and exposure times
for each galaxy.

\par
Data reduction for the optical images was carried out using the 
IRAF (\cite{tod86}) and VISTA (\cite{sto88}) packages.
The images were bias--subtracted and flatfielded with twilight or
dome flats for each filter.  An apparent gradient in the background of 
some of the $V$, $R$, and $I$ frames taken at Palomar remained after these
steps. A planar
surface was fit to these frames after masking out the galaxy, then subtracted
before further processing.  Cosmic rays were identified with the 
IRAF ``cosmicrays'' package and were replaced with an average of surrounding
pixel values.  If more than one data frame was available for a galaxy in
a specific filter, the frames were registered using the coordinates of
several bright stars, scaled to the same exposure time, and
a zeropoint offset was applied to match the background levels.
The frames were then averaged (for two) or median combined (for 
three or more).  Some of the images are saturated at the galaxy centers; no 
correction is applied but the saturated pixels were flagged for future 
consideration.  Sky levels were determined in boxes away from the galaxy. 
In some cases (flagged in Table~\ref{galxd2}) the galaxy nearly fills the
frame and thus the background
level is somewhat uncertain.  On images in which several regions of sky away 
from the galaxy can be measured, the typical r.m.s. scatter in sky levels
across the frame is less than 2\% of the sky value.

\subsection{Data Calibration}

\par
Calibrations were applied to each image to obtain the approximate
relative alignment of the multicolor light profiles presented in Section~3.
The UIT data were calibrated using standard stars measured by
the International Ultraviolet Explorer ($IUE$).  The resulting flux calibration
is estimated to be accurate to $\sim~15\%$ (\cite{ste97}); this value 
includes the uncertainty in the final $IUE$ calibration.  Magnitudes are
on the monochromatic system:
\begin{equation}
{\rm mag_{\lambda}} \: = \: -2.5 {\rm log_{10}} (f_{\lambda})\; - \; 21.1
\end{equation}
where the flux is in units of erg cm$^{-2}$s$^{-1}$ $\AA^{-1}$.
Calibration data for each image are stored in the image headers of the UIT
data products.

\par
Because many of the optical data were obtained under non--photometric 
conditions, the frames were calibrated via comparison to published aperture
photometry for each galaxy.  We used the catalog of \cite{pru98}, excluding
any data whose zeropoint was found by these authors to be systematically 
offset by more than 0.05 mag. from that galaxy's curve of growth.
We calculated the difference between a synthetic aperture
magnitude measured on our galaxy image and the published value for
each of the tabulated apertures that fit on our frame.  After examining
the residuals as a function of ($B-V$) or ($V-I$) color for those galaxies 
with sufficient
published data over a range of colors, we found no strong evidence for 
color terms in the calibration. Therefore, we simply used the mean difference
between the measured and published magnitudes as a calibration constant.
For images in which the galaxy center was saturated, we compared the
observed and published magnitudes in annuli between the smallest published
aperture that enclosed all the saturated pixels and subsequent apertures.
The typical scatter around this mean difference ranges from 0.02--0.15 mag.,
which is not unexpected for a mix of photometric and non--photometric data from
several different instruments.  The uncertainty in optical aperture magnitudes
due to sky and instrument noise is typically $\leq 3\%$.  If no catalog
data were available for a particular galaxy/filter, we substituted an average
of the calibration constants determined in that filter for other galaxies
observed during the same night.  As all but one of these cases 
involved calibration of $R$ or $I$--band data, airmass differences between 
the galaxies that were bootstrapped and those for which calibrations were
determined from the literature are likely to contribute only small errors.
Galaxies for which this procedure was necessary are flagged in 
Table~\ref{galxd2}; their calibrations are
correspondingly less certain than the others.  Overall, combining the 
noise and calibration errors, we estimate an uncertainty in 
the optical magnitudes of $\leq 15\%$, with the exact value for each
galaxy depending on the quality of the observations and the availability
of published data.

\subsection{Image Alignment and Stellar Masks}
\par
In order to facilitate the direct comparison of galaxy morphology at UV and 
optical wavelengths, we transformed the coordinates of the
optical data onto the system of the UIT data and smoothed all data for a
particular galaxy to the same resolution.  Because the FUV images contain
few or no stars, the coordinates of UV--bright star--forming regions were
used to determine a linear transformation (rotation, scaling, and
translation) between the shortest wavelength optical image and
the UIT data.  After the transformation was applied to that optical image, 
the centroids of several bright stars were used to align additional
optical data.  Finally, we measured the FWHM of the point spread function (PSF)
on each optical image and smoothed the frame
to the UIT resolution of 3\arcsec.  The PSF FWHM for two galaxies, M51 and 
NGC 925, were slightly larger than the UIT FWHM (4$\arcsec$ and 4.5$\arcsec$ 
respectively); in these cases the 
UIT images were smoothed to the optical resolution.
The final images have the UIT scale of 1.136\arcsec/pixel, implying
physical scales ranging from $\sim 10 - 140$ pc/pixel  for our sample galaxies.

\par
Foreground stars and any remaining artifacts were masked out on the optical
and FUV data before producing the images and surface brightness 
profiles presented in Section 3.  For the optical data, we identified
stars on the longest wavelength image and created a mask that
was used to interpolate over those pixels on all of the optical images.
Most of the sample galaxies lie at high Galactic latitude, so few foreground
bright stars were present.  However, in the case of NGC 2403,
badly saturated bright 
stars in the $R$ and $I$--bands necessitated masking large areas of those
images.  For this galaxy, we created a separate mask for the $UBV$ images, 
which did not require such extensive corrections.
In a few cases it was difficult to determine whether bright spots within
the galaxy were foreground stars or HII regions; 
here we carefully compared the optical and FUV images and masked only those
spots that were not visible in the FUV. Some galaxies also required 
a FUV mask to interpolate over small, bright point--like or streak artifacts
 (mostly due to cosmic rays) that remained after prior data processing. 

\section{Images and Surface Brightness Profiles}

\par
In Figures~\ref{ims}($a-n$), we present multiwavelength images for the
sample galaxies.  The sky backgrounds have been removed, foreground stars 
have been masked as described above. For display purposes only, image 
pixel values have been converted to ``calibrated'' flux units such that: 
\begin{equation}
{\rm mag \; arcsec^{-2}} \: = \: 26. \; - \; 2.5 {\rm log_{10}(pixel \; value)}
\end{equation}
The images for a given galaxy are displayed using the same range of calibrated 
brightness for each filter, so that pixels with the same brightness
(flux/$\AA$) have a constant gray level on each image.
This mode of presentation shows the relative limiting surface brightnesses
of the data at different wavelengths.  To conserve space,
we do not show every optical 
image, as those from adjacent optical filters often look nearly identical.  
It is quite difficult to simultaneously
show structure in the bright centers of galaxies and faint outlying
features, so we present additional images of some galaxies 
with different scales or normalizations in Figures~\ref{vfigsf} and
\ref{vfigearly}.
The four FUV--bright galaxies shown in Figure~\ref{vfigsf} are displayed with
a much larger scale to show the central regions.
In Figure~\ref{vfigearly}, we show several galaxies with red FUV--optical
colors, mainly early--type galaxies or those with a prominent optical bulge.
Here we show the optical images with a much larger stretch
than the FUV data, so that the central structure of the optical image can
be compared to the FUV.  We also plan to make our images available to the
community in digital form via the NASA/IPAC Extragalactic Database (NED);  
they can then be renormalized as appropriate for any application.

\par
We extract surface brightness profiles for each galaxy by azimuthally
averaging around concentric ellipses.  The ellipse center, ellipticity, 
and position angle (P.A.) for each galaxy were determined by running an 
ellipse--fitting package in VISTA (based on the method of \cite{ken83})
on the longest available wavelength image, usually $R$ or $I$. 
The centroid was determined 
in a small box around the visual center of the galaxy, taking care to use
an image in which the center was not saturated.  The ellipticity and P.A.
 are determined from the outer ellipses, where their values were typically
stable over a large range in radii.  These ellipse parameters were used to
obtain surface brightness profiles at all wavelengths, in order to avoid 
problems with determining centroids and fitting ellipses on the
often asymmetric and irregular FUV images.  For one galaxy, NGC 3115, we
were unable to align the optical and FUV images due to a lack of bright 
features or foreground stars in the FUV.  We determined a separate FUV 
ellipticity and P.A. for this galaxy using the ellipse fitting routine. 
The ellipticity and P.A. for each galaxy, which are given in 
Table~\ref{galxdat},
are quite similar to the $RC3$ (\cite{dev91}) values 
($\Delta \epsilon \sim 0.05$ and $\Delta {\rm P.A.} \leq 15^{\circ} $).
The surface brightness profiles derived for each sample galaxy are shown in 
Figure~\ref{sbprofs}($a-g$), with data from the various filters denoted
by different symbols.  Images in which the galaxy center was saturated are
identifiable by their flat profiles in the central regions; this is the
case for several optical images and the inner $\sim$10$\arcsec$ of
the FUV data for NGC 4151.
Profiles for a subset of our sample were compared to those extracted
by Fanelli et al. (1997b) from the same data; no significant differences
are noted.
The profiles have not been corrected for foreground Galactic extinction. 
However, the foreground $A_{B}$ values (given in Table~1) are generally small, 
so even FUV extinctions of $\sim 2\times A_{B}$ amount to $\leq 0.5$mag and
would have little effect on the plots of Figure~\ref{sbprofs}.

\par
Three systems in our sample are interacting pairs for which it is difficult
to separate the light profiles of the two galaxies:  NGC 4038/9, NGC 3226/7, 
and NGC 5194/5 (M51a,b).  (NGC 4647 and NGC 4649 are a close pair that may be
undergoing a mild interaction, but their FUV isophotes do not overlap in our
images.)  For NGC 5194/5, the companion galaxy NGC 5195 is 
undetected in the FUV image and thus is not analyzed separately.  The 
ellipse parameters for NGC 5194 are determined from a region unaffected by
the companion on the optical images.  Many of the outer ellipses cut through
the companion, creating a slight ``bump'' in the outer part of the 
optical surface brightness profiles.  The outer ellipses  for NGC 3227
encompass all of the FUV light and a significant fraction of the optical 
light of its companion NGC 3226.  Within the available software packages, 
it is impossible to avoid this situation 
without either severely truncating the NGC 3227 profile or changing the
ellipticity so much that it no longer accurately reflects the shape of 
NGC 3227.  However,  the NGC 3226 profile of the region with
detectable FUV emission is relatively unaffected.  The merger system
NGC 4038/9 is treated as one galaxy, as the FUV image does not suggest a
simple division into two separate disks.  We created an image of
this system
consisting of only pixels with flux greater than 1.5$\sigma_{\rm sky}$ above
the background that also have at least 4 adjacent pixels above the threshold.
(``Adjacent'' pixels are defined as those that share either an edge
or corner with the pixel in question; thus 
each pixel has 8 adjacent neighbors.)  This image is approximately circular, 
so we defined an aperture center at the geometric center of the FUV 
image and used circular annuli to measure the light profiles.

\par
Color profiles are produced by subtracting the surface brightness
profiles (in mag~arcsec$^{-2}$) of different filters.
The use of a single set of ellipse parameters
for all images of a particular galaxy ensures that the light profiles at
each wavelength sample the same physical region and thus can be 
compared in this simple manner.  Color profiles for many of the sample galaxies
will be presented and discussed in the next section.

\section{Optical and FUV Morphology}

In this section we present a qualitative discussion of the differences in 
apparent morphology between FUV and optical images.  The morphological 
types included in each subsection below are approximate, stemming in part
from the fact that several galaxies
have characteristics associated with more than one ``type''.  Rather
than discussing each galaxy individually (as is done in Marcum et al. 2000),
we summarize the 
changes in apparent morphology with wavelength for various Hubble types.
(Interested readers can see the results for each individual galaxy from
the data presented in Figure~\ref{ims}.) Throughout
this section, we also consider implications of our analysis for the study
of high--redshift galaxies that are observed in their rest--frame UV.
It is important to remember that this level of discussion assumes
no evolution of
the star--formation rate: we simply consider what galaxies {\it identical} to 
local ellipticals, spirals, and irregulars would look like at high redshift.
Galaxies undergoing their initial starburst or a subsequent period
of increased star formation activity due to mergers, etc. would, of course,
look different from the quiescent ellipticals and ``normal'' spirals
in this sample and may more
closely resemble star--forming irregular or peculiar systems.

\subsection{Elliptical and S0 Galaxies}

\par
The elliptical and S0 galaxies in our sample feature smooth, compact FUV
emission centered on the nuclear regions (see also the detailed analysis
of UIT data
for E/S0 galaxies in Ohl et al. 1998).  In galaxies of this type,
FUV emission is produced not by young stars but by the hot, evolved
population responsible for the spectrosopic UV upturn (\cite{oco99}).
The FUV light profiles of 
these systems -- NGC 3379, NGC 3384, NGC 4649, NGC 3226, and NGC 3115 -- 
drop off much more rapidly than the optical light, and the colors throughout
are quite red (FUV--$R$ $\sim$ 2--5).  FUV half--light radii for these 
galaxies are $\sim$20--35\% of their $B$--band values tabulated in 
the $RC3$ (we cannot compare directly to our optical data because most
of the E/SO optical images suffer saturation in the center).
The only exception to this pattern is
NGC 1510, the interacting companion of NGC 1512.  The FUV emission in this
galaxy is comparable to the optical light in both
brightness and physical extent and is likely due to tidally induced star 
formation rather than the evolved population.  Although it is classified in the 
$RC3$ as a peculiar S0 galaxy, \cite{san79} classify NGC 1510 as an
Amorphous galaxy, and it is also noted to be a star--forming blue compact
dwarf (BCD) in
\cite{kin93} and an HII galaxy in NED.
(The other S0/Amorphous galaxy in our sample, NGC 5195, is also interacting
with companion galaxy NGC 5194 but is not detected in the FUV.)
\par
It is difficult to predict the appearance of high--redshift E/S0 galaxies in
their rest--frame FUV because the present--day UV emission mechanism, that of 
evolved stars, is likely to be weak to nonexistent at redshifts approaching 
$z \sim 1$ (Brown et al. 1998, 2000).  Indeed, the UV upturn phenomenon is a
potentially sensitive age indicator for evolved stellar populations, and
thus its evolution with redshift could help identify the formation epoch of
early--type galaxies (\cite{yi99}).  At high redshifts ellipticals may be 
undergoing an initial starburst and/or the UV upturn population will not yet
have had time to evolve.  In this case, the progenitors of present--day
elliptical and S0 galaxies may look like local starbursts or mergers
depending on their formation mechanism and the timing of the observations.
However, if some high--redshift E/S0's were quiescent and identical to local 
early--type galaxies ({\it i.e.} assuming a very high formation redshift),
they would still have the smooth and symmetric light distributions associated
with early types.  Their FUV emission would be much more compact than the
optical light, which would lead to difficulties in detection and confusion in
estimating their typical size.  

\subsection{Early--Type Spiral Galaxies}

\par
Galaxies of Hubble type Sa--Sbc tend to appear as later--type
spirals in the FUV than at optical wavelengths, primarily
due to fading of the red
bulge and bar in UV images. A significant fraction of the FUV light in 
these systems comes from star--forming regions in the spiral arms, which 
are visible in the images of Figure~\ref{ims} and appear as ``bumps''
in the FUV surface brightness profiles of Figure~\ref{sbprofs} (see also 
\cite{fan97b}).
Some early--type spirals, such as M63 and NGC 1672, also appear more
patchy or fragmented in the FUV because old stars in the underlying smooth
disk do not produce much light at very short wavelengths.  
These results are in agreement with earlier work by O'Connell (1997a) and
O'Connell \& Marcum (1996) for smaller samples of galaxies.
The dramatic fading of 
light in the central regions (also noted by \cite{oco97a} and \cite{wal97b})
produces a global pattern in which the central FUV--optical colors are much
redder than the outer regions, as shown in Figure~\ref{ecolpr} for M~51, 
NGC~1566, and M~63.   
Two notable exceptions to this pattern are NGC 4151, in which the Seyfert 1 
nucleus dominates the central FUV emission and is much brighter than the
weak spiral arms, and NGC 1068, in which the FUV--bright Seyfert 2 nucleus
and surrounding star--formation combine to produce a very blue
center.  The UV morphologies and luminosities of the AGN and star--forming
components in these two galaxies are discussed in more detail by \cite{fan97a}.

\par
Many early--type spirals also show evidence of star--forming inner rings or
circumnuclear star formation, which often dominate the FUV images and light
profiles.  Images of these features, which are present in NGC 1097, NGC
1512, NGC 1672, NGC 3310, NGC 3351, and NGC 4736, are shown in 
Figure~\ref{sfring}, and the associated color profiles are plotted in 
Figure~\ref{ringcolpr}.  The  star--forming rings appear as spikes of FUV
brightness or very blue color in the radial profiles of Figures~\ref{sbprofs}
and \ref{ringcolpr}.
Inner and circumnuclear rings are thought to be orchestrated by dynamical 
resonances, which are often associated with central bars 
({\it e.g.} \cite{sto96}; \cite{but96}).  The FUV properties of some
specific ringed galaxies are discussed by Waller et al. (1997c, 2000)
for NGC 4736, Smith et al. (1996) for NGC 3310, and Marcum et al. (2000).

\par
Galaxies classified from their optical images as early--type spirals may present
a variety of different appearances when viewed at high redshift, in their
rest--frame FUV.  Optically barred galaxies such as NGC 1097 and NGC 1365 will 
appear unbarred in the ultraviolet. 
Isolated galaxies such as NGC 1672 that appear highly 
fragmented in the FUV may be mistaken for an ongoing merger.
Images of distant systems may recover only the FUV--bright star--forming
rings or circumnuclear star formation, yielding little evidence of
the underlying regular structure seen in the galaxy's optical light (see
also Waller et al. 1997b).
On the other hand, galaxies such as NGC 4151 appear to have an earlier
morphological type in the FUV due to the prominence of
highly symmetric and strongly peaked
emission from the active nucleus.  Finally, the appearance of a galaxy
undergoing a global starburst,
such as NGC 3310, does not differ a great deal between the FUV and optical
because the light at both wavelengths is dominated by young stars.  
We have found that all of these effects can vary in magnitude,
such that some galaxies would be assigned nearly the same Hubble type in
the optical and FUV, while others would have very different classifications.
Overall, most optically--classified early--type spirals present the 
appearance of a later type in the FUV, but a small fraction may appear
to have identical or earlier FUV types in the presence of starburst activity
or an AGN.

\subsection{Late--Type Spirals, Irregulars, and Starbursts}

\par
Galaxies that are optically classified as late--type spirals generally do
not show as dramatic a difference between their FUV and optical morphologies as
do early--type spirals.  Although they often appear
somewhat more patchy in the FUV, there is no prominent optical bulge whose
absence at short wavelengths is noticeable.  Star formation
occurs over a large fraction of the galaxy, producing both FUV and optical
light in the same physical regions.
A number of these galaxies, including NGC 4449, NGC 925, NGC 1313, and IC 2574,
show color profiles that have no definite trend with radius.  These 
profiles, examples of which are displayed in Figure~\ref{flatcolpr}, 
are either flat or bumpy depending on the relative prominence of star--forming 
regions against the background disk.  
Figure~\ref{bluecolpr} shows color profiles for the central starburst galaxies
M83, NGC 4214, and NGC 5253, which  are bluest in their centers. 
NGC 4214 and NGC 5253 have global color gradients in the
opposite direction of that observed in early--type spirals due to the
presence of a blue starburst superposed on a more extended old stellar
distribution.  This ``starburst core'' morphology has been previously
noted in FUV--optical image comparisons of NGC 4214 (Fanelli et al. 1997c)
and NGC 5253 (Kinney et al. 1993). A
central starburst combined with bright star--forming regions in the outer
disk can also produce a flat color profile, which is the case for NGC 2903 
(also shown in Figure~\ref{bluecolpr}). 

\par

We find that
the morphological $k$--correction for late--type galaxies is expected
to be highly dependent on whether or not they host starburst activity.
As evident from their similar FUV and optical morphology, non--starbursting
late--type galaxies (dominated by ongoing disk star formation)
viewed at high redshift would not generally appear to have a very different
type from their local counterparts.  Increased patchiness in the FUV (if
it were still detectable at large distances) might suggest a slightly later
type than the optical appearance.  However, galaxies with central starbursts
like those
seen in NGC 4214 and NGC 5253 might present compact cores surrounded by 
diffuse light, similar to some objects seen on deep HST images
(Giavalisco, Steidel, \& Duccio Macchetto 1996a). 
The symmetric central burst in NGC 5253
might even appear to be an E or S0 galaxy in the absence of the old stars.
If starburst systems are more prevalent at high redshift, their strong
influence on the rest--frame UV morphology must be considered when
studying the evolution of the galaxy population.

\subsection{Interacting Systems}

\par
As might be expected, the morphological $k$--correction for interacting
galaxies depends strongly on the details of the interaction itself.
Three objects in our sample are currently undergoing interactions of various
degrees:  NGC 4038/9 (the Antennae), NGC 3226/7, and NGC 5194/5 (M51).
The merger of two disk galaxies in NGC 4038/9 has produced a significant
amount of massive star formation, concentrated mainly between the two
disks and around the edge of the NGC 4038 disk.  Thus the FUV image shows
a partial outline of the merger system, and, without the corresponding optical
data, suggests the appearance of a single, drastically perturbed galaxy.
However, the old populations visible in long--wavelength optical images
clearly delineate the inner disks and tidal tails that identify NGC 4038/9
as an ongoing disk--disk merger.  In spite 
of the dramatic difference between its FUV and optical appearances,
NGC 4038/9 is clearly peculiar at both wavelengths and thus would
not be misclassified as any sort of ``normal'' Hubble type galaxy when
viewed at high redshift.
Both galaxies in the NGC 3226/7 interacting pair present FUV morphologies
consistent with what would be expected from their Hubble types:
the elliptical NGC 3226 appears highly
concentrated and the Sa NGC 3227 shows emission from the center
and a few faint disk star--formation regions.  In this case, the influence
of the interaction in determining the FUV morphology is confined  to
its possible role in driving the Seyfert nucleus and circumnuclear
star formation in NGC 3227 (\cite{kee96}; \cite{gon97}).  
NGC 5195 is not detected in the
FUV, likely due to obscuration by dust from a foreground spiral arm in 
NGC 5194 as well as possible internal extinction (see \cite{san61}.
\cite{bar98} also suspect that dust is responsible for their failure to
detect NGC 5195 at 2200$\AA$.)  Observed at high redshift in its
rest--frame UV, the M51 double--galaxy system would appear to be an
isolated spiral galaxy.  Although the examples discussed above demonstrate the 
difficulty of relating UV and optical morphology for interacting
systems, interactions that are significant enough to severely distort
the galaxies involved are often associated with enhanced star formation rates
that may identify the system as ``abnormal'' in the FUV as well
(\cite{ken98} and references therein).

\subsection{Dust and Morphology}

\par
Throughout this section we have discussed FUV morphology mainly in terms
of the current star formation, but, as suggested by the example of M51,
one cannot neglect the
possibility that extinction also influences the UV appearance
of galaxies.  It is difficult to ascertain whether the 
``patchiness'' or fragmented appearance of many galaxies in the FUV is due
to internal extinction or to the pattern of massive star formation.
Two of the most highly inclined ($i \sim 65^\circ$) galaxies 
in our sample, NGC 2841 and NGC 2903, 
have asymmetric FUV emission with the bright side corresponding to what
appears to be the unobscured side of the optical disk.  On the other hand, 
NGC 2403 and IC 2574 also have $i \geq 60^\circ$ but do not 
appear to suffer heavy extinction on one side of the disk.  These results
provide further confirmation of the variations in FUV brightness of edge--on
galaxies noted by Smith et al. (1997, 2000) for 
NGC~4631 (FUV--bright) and NGC~891 (undetected).  An extreme case of
extinction effects was discussed above for NGC 5195 (M51b), which is completely
obscured in the FUV by the foreground spiral arm of NGC 5194 (M51a).  
As noted in the introduction, the role of dust in determining the observed
FUV morphology depends strongly on the dust--star geometry of individual
galaxies and is thus difficult to predict from optical images alone.

\section{Summary and Discussion}

\par
We have presented FUV and optical images and surface brightness profiles for
34 galaxies observed with the UIT and compared the apparent morphology at
different wavelengths in the context of interpreting images of high--redshift
objects. The UIT data represent the most detailed set of FUV images obtained
to date for large, nearby galaxies that are well--resolved. 
Present--day elliptical and S0 galaxies appear smooth and symmetric in FUV as
well as in optical images, but they are much more compact in the FUV and
may be faint and/or unresolved at high redshift unless they are 
actively forming stars at that particular look--back time.
The majority of spiral galaxies in our sample
appear to have a later Hubble type in the FUV than at optical wavelengths
due to increased patchiness in combination with the fading of light from
the bulge and/or bar populations.
This effect is particularly dramatic for the early--type spirals, in which
bulges and often bars are prominent optical features.  The optical and FUV
morphologies of late--type spirals and irregular galaxies do not differ as much 
as the earlier types because young stars dominate the light in both spectral
regimes.  Some galaxies appear highly fragmented in the FUV images in spite of
a regular optical morphology, making it 
hard to determine from UV data alone whether they are single systems or 
multiple mergers.
Even assuming their current rates and patterns of star formation, many spirals
viewed at high redshift would be assigned later Hubble types than their
local counterparts due to the bandshifting effects.
Central or circumnuclear starbursts and star--forming rings, and bright
spiral arms dominate the FUV light and FUV--optical color profiles of many
galaxies.  Depending on the relative geometry of dust and young stars, 
extinction can play a varying role (small to dominant) in
determining the apparent
FUV morphology.  From the galaxy sample presented here, it is apparent that
understanding the correlation between FUV and optical morphology depends on
one's knowledge of not only the global optical Hubble type but also 
of smaller scale features, mostly those due to recent star--formation patterns.

\par
The fragmented appearance of many of the FUV images and the prominence of
young stars in structures such as rings or fragments of spiral arms 
produces FUV morphologies that do not readily fit into the traditional bins
of the Hubble sequence.  In order to characterize
rest--frame UV galaxy populations at both low and high redshift, it will be
necessary to find new ways of describing the morphology that do not by default
cause the majority of galaxies to fall into the ``peculiar'' or other 
catch--all bins.  It would also be desirable to have an objective 
classification scheme to express the similarities or differences between 
local and distant galaxies in quantitative terms.  However, UV data pose
problems for many automated classification schemes because galaxy centers
are often ill--defined and the patchy nature complicates the selection of
an aperture that encloses a single system.  Future work in this area
will use UIT data to identify new features or indices that better describe
the UV morphology of galaxies. 

\par
The dataset presented in this paper will be valuable for studying the
morphological $k$--correction that must be taken into account when
interpreting deep optical images such as the {\it Hubble Deep Field} 
(HDF, \cite{wil96}).  The dependence of FUV morphology on features that are
considered minor or secondary details in the optical highlights the
difficulty of studying
evolution in the galaxy population by comparing rest--frame UV observations
of high--redshift galaxies to the optical properties of local 
samples.  The comparison between local and distant UV galaxy 
images is much more direct than relying on estimates of the UV morphology 
extrapolated from optical light and colors ({\it e.g.} Abraham et al. 1997).   
Indeed, a small sample of UIT galaxy images from the {\it Astro--1} 
mission have
been used to simulate the appearance of high--redshift counterparts of 
nearby bright galaxies (Giavalisco et al. 1996b).
In a subsequent paper (\cite{kuc00}), we simulate the cosmological distance
effects of surface brightness dimming and loss of spatial resolution on 
the FUV images of our larger {\it Astro--2} galaxy sample.
These simulations provide examples of how local 
galaxies would appear in the HDF at redshifts of $\sim 1 - 4$.  In that paper, 
we also attempt to quantify the effects of bandshifting and distance on 
morphology using simple parameters such as the central concentration and
asymmetry of galaxies.

\par
One of the primary goals of studying morphology is to infer the past
or ongoing physical processes that have shaped galaxies.
In combination with current knowledge about the dynamical states
of nearby galaxies, our multi--wavelength dataset and future UV imaging
data from the GALEX sky survey (\cite{mtin97}) could be used to relate
UV morphology to physical structures or properties.  For example, 
Waller et al. (1997b, 2000) find that UV--bright starburst rings likely
result from dynamical resonances with a bar component. Thus ring features
may predict the presence of an underlying disk and bar system even in 
UV images that appear to be fragmented or otherwise dynamically disorganized.
It would also be of interest to further investigate the suggestion of 
Waller et al. (1997b) that optical Hubble type may be correlated with 
FUV--$V$ color (although rest--frame $V$--band data will not necessarily
be available for high--redshift galaxies).
If relations between UV morphology and physical properties can be found in 
local galaxies {\it and} shown to extend to large look--back times,
they could provide a valuable tool for studying the distant universe
through rest--frame UV imaging.
\acknowledgements
The authors gratefully acknowledge the help of R. Bernstein, J. Parker, R. Phelps, and N. Silbermann for obtaining some of the optical imaging data presented
here.  We thank O. Pevunova for assistance with preliminary data reduction.
We are also grateful to the anonymous referee for many constructive suggestions.
Funding for the UIT project was provided through the Spacelab office at NASA 
Headquarters under Project Number 440-51.  WLF and BFM acknowledge support from
the Astro--2 Guest Investigator Program through grant number NAG8--1051.
Some of the data presented here were
obtained at CTIO, which is operated by AURA as part of NOAO under a cooperative 
agreement with the National Science Foundation.  This research has made use of
the NASA/IPAC Extragalactic Database (NED), which is operated by the Jet 
Propulsion Laboratory, California Institute of Technology, under contract
with NASA.

\figcaption{FUV and optical images of sample galaxies.  All images are registered to the FUV coordinate system and oriented with North up and East to the left.
As noted in the text, the images are displayed such that pixels of the same
brightness (flux) have constant
gray level for all images of the same galaxy.  Scale bars, which are relevant
to all images of a particular galaxy, are shown on each of the FUV panels. \label{ims}} 
\figcaption{FUV and optical images of FUV--bright galaxies rescaled to show
bright central regions.  As in Figure~\ref{ims}, pixels of the
same brightness have constant gray level on all images of a galaxy. \label{vfigsf}}
\figcaption{FUV and optical images of galaxies with red FUV--$V$ colors, 
renormalized so that optical data are not badly saturated in the display.
Here, the optical data are displayed with a larger stretch than the FUV images
in order to compare the central structure in both images. \label{vfigearly}}
\figcaption{FUV and optical surface brightness profiles of sample galaxies. The surface brightnesses are calibrated as described in the text, then offset by a
constant to better show the relative shapes.  Data from the optical filters are
offset for ease of viewing; offsets are noted on each panel.  Symbols for the 
different filters are as follows: FUV=filled diamonds, $U$=open circles,
$B$=plus symbols, $V$=open triangles, $R$=open asterisks, 
$I$=open squares. \label{sbprofs}}
\figcaption{FUV--optical color profiles for early--type spiral galaxies. The
galaxy name and constant offset (where applicable)  are noted next to each profile. \label{ecolpr}}
\figcaption{FUV and optical images of circumnuclear and inner star--forming rings in early--type spirals.  The images are aligned, oriented, and scaled as
described in Figure~\ref{ims}. \label{sfring}}
\figcaption{Color profiles for early--type spirals with star--forming rings or
circumnuclear star--formation.  As in Figure~\ref{ecolpr}, the galaxy name and constant offset are noted next to each profile.  Locations of the rings and 
circumnuclear star--formation (CSF) regions are labeled for each galaxy. \label{ringcolpr}}
\figcaption{Color profiles for late--type galaxies with flat FUV--optical color
gradients.  As in Figure~\ref{ecolpr}, the galaxy name and constant offset are
noted next to each profile. \label{flatcolpr}}
\figcaption{Color profiles for late--type galaxies with central starbursts.
As in Figure~\ref{ecolpr}, the galaxy name and constant offset are
noted next to each profile. \label{bluecolpr}}

\onecolumn
\pagestyle{empty}
\small
\begin{deluxetable}{lllccrrrr}
\tablenum{1}
\tablewidth{0pt}
\tablecaption{Basic Data for UIT Sample Galaxies \label{galxdat}}
\tablehead{
\colhead{Name} 
& \colhead{$\alpha $ (2000)\tablenotemark{a}}
& \colhead{$\delta $ (2000)\tablenotemark{a}}
& \colhead{Type\tablenotemark{b}}
& \colhead{Activity\tablenotemark{b}}
& \colhead{D\tablenotemark{c}}
& \colhead{$A_{B}$\tablenotemark{c}}
& \colhead {$\epsilon$\tablenotemark{e}}
& \colhead {P. A.\tablenotemark{f}} 
}
\startdata
NGC 925 & 02 27 16.8 & +33 34 41 & SAB(s)d &  \nodata & 9.4 & 0.25 & 0.40 & 115 \nl
NGC 1068 (M 77) & 02 42 40.2 & $-$00 00 48 & RSA(rs)b &  Sy 2 & 14.4 & 0.05 & 0.20 & 84 \nl
NGC 1097 & 02 46 18.9 & $-$30 16 21 & SB(s)b &  Sy 1 & 14.5 & 0.07 & 0.32 & 140 \nl
NGC 1313 & 03 18 15.5 & $-$66 29 51 & SB(s)d &  \nodata & 3.7 & 0.03 & 0.20 & 40 \nl
NGC 1365 & 03 33 36.6 & $-$36 08 17 & SB(s)b &  Sy 1.8 & 16.9 & 0.00 & 0.45 & 32 \nl 
NGC 1510 & 04 03 32.6 & $-$43 24 01 & S0,pec &  H II & 10.3 & 0.00 & 0.12 & 115 \nl
NGC 1512 & 04 03 54.6 & $-$43 21 03 & SB(r)a &  \nodata & 9.5 & 0.00 & 0.36 & 56 \nl
NGC 1566 & 04 20 00.4 & $-$54 56 18 & SAB(s)bc  & Sy 1 & 13.4 & 0.00 &  0.21 & 40 \nl
NGC 1672 & 04 45 42.2 & $-$59 14 57 & SB(s)b &  Sy 2 & 14.5 & 0.00 & 0.13 & 161 \nl
NGC 2403 & 07 36 54.5 & +65 35 58 & SAB(s)cd &  \nodata & 4.2 & 0.16 & 0.44 & 115 \nl
NGC 2841 & 09 22 01.8 & +50 58 31 & SA(r)b &  LINER, Sy 1 & 12.0 & 0.00 &  0.56 & 147 \nl
NGC 2903 & 09 32 09.7 & +21 30 02 & SAB(rs)bc &  H II & 6.3 & 0.07 &  0.53 & 24 \nl
NGC 3115 & 10 05 14.1 & $-$07 43 07 & S0 &  \nodata & 6.7 & 0.10 &  0.66\tablenotemark{g} & 43\tablenotemark{f} \nl
NGC 3226 & 10 23 27.4 & +19 53 55 & E2*,pec &  LINER & 23.4 & 0.02 &  0.20 & 10 \nl
NGC 3227 & 10 23 31.5 & +19 51 48 & SAB(s)a,pec &  Sy 1.5 & 20.6 & 0.02 & 0.55 & 157 \nl
IC 2574 & 10 28 22.5 & +68 24 39 & SAB(s)m &  \nodata & 2.7 & 0.06 & 0.59 & 62 \nl
NGC 3310 & 10 38 46.1 & +53 30 08 & SAB(r)bc,pec &  H II & 18.7 & 0.00 & 0.22 & 170 \nl 
NGC 3351 (M 95) & 10 43 58.0 & +11 42 15 & SB(r)b &  H II & 8.1 & 0.04 & 0.32 & 17 \nl 
NGC 3379 (M 105) & 10 47 49.9 & +12 34 57 & E1 &  \nodata & 8.1 & 0.05 &  0.13 & 65 \nl
NGC 3384 & 10 48 17.2 & +12 37 49 & SB0* &  \nodata & 8.1 & 0.05 & 0.50 & 51 \nl
NGC 3389 & 10 48 27.9 & +12 32 01 & SA(s)c &  \nodata & 22.5 & 0.06 & 0.55 & 108 \nl
NGC 4038 & 12 01 52.9 & $-$18 51 54 & SB(s)m,pec & \nodata & 25.5 & 0.05 &   0.00\tablenotemark{h} & \nodata \nl
NGC 4039 & 12 01 53.9 & $-$18 53 06 & SA(s)m,pec & \nodata & 25.3 & 0.05 &  0.00\tablenotemark{h} & \nodata \nl
NGC 4151 & 12 10 33.0 & +39 24 28 & SAB(rs)ab*,Pec & Sy 1.5 & 20.3 & 0.00 & 0.36 & 145 \nl
NGC 4156 & 12 10 49.7 & +39 28 24 & SB(rs)b &  LINER & 90.4\tablenotemark{i} & 0.00 &  0.20 & 45 \nl
NGC 4214 & 12 15 39.5 & +36 19 39 & IAB(s)m &  H II & 3.5 & 0.00 & 0.18 & 132 \nl
NGC 4449 & 12 28 11.4 & +44 05 40 & IBm &  \nodata & 3.0 & 0.00 & 0.46 & 60 \nl
NGC 4647 & 12 43 32.4 & +11 34 56 & SAB(rs)c & \nodata & 16.8 & 0.04 & 0.20 & 135 \nl
NGC 4649 (M 60) & 12 43 40.3 & +11 32 58 & E2 & \nodata & 16.8 & 0.04 & 0.17 & 130 \nl
NGC 4736 (M 94) & 12 50 53.6 & +41 07 10 & (R)SA(r)ab &  LINER & 4.3 & 0.00 & 0.22 & 95 \nl
NGC 5055 (M 63) & 13 15 49.3 & +42 02 06 & SA(rs)bc & H II, LINER & 7.2 & 0.00 & 0.47 & 102 \nl
NGC 5194 (M 51) & 13 29 53.3 & +47 11 48 & SA(s)bc,pec &  Sy 2.5 & 7.7 & 0.00 & 0.30 & 30 \nl
NGC 5236 (M 83) & 13 37 00.3 & $-$29 52 04 & SAB(s)c &  H II & 4.7 & 0.14 & 0.10 & 80 \nl
NGC 5253 & 13 39 55.9 & $-$31 38 41 & Pec &  H II & 3.2 & 0.19 & 0.57 & 43 \nl
NGC 5457 (M 101) & 14 03 12.5 & +54 20 55 & SAB(rs)cd &  \nodata & 5.4 & 0.00 &  0.00 & \nodata \nl 
\vspace*{-0.15in}
\tablenotetext{a}{Data from the $RC3$.}
\tablenotetext{b}{Nuclear activity classification, from the NASA Extragalactic Database (NED).}
\tablenotetext{c}{Distances in Mpc from Tully 1988 ($H_{0} = 75$ km/s/Mpc), except as noted.}
\tablenotetext{d}{Foreground Galactic extinction values from Burstein \& Heiles 1984.}
\tablenotetext{e}{Adopted ellipticity for azimuthal averaging in surface brightness profile calculations, defined as $\epsilon = 1 - \frac{b}{a}$.}
\tablenotetext{f}{Adopted major axis position angle for azimuthal averaging in surface brightness profile calculations.}
\tablenotetext{g}{An ellipticity of 0.29 and a P. A. of 118 were used for the
FUV image of NGC 3115, which was never registered to the optical data (see Section 3).}
\tablenotetext{h}{NGC 4038/9 is treated as one galaxy for profile calculation, see Section 3.}
\tablenotetext{i}{Distance from V$_{\rm GSR}$, $H_{0} = 75$ km/s/Mpc.}
\enddata
\end{deluxetable}

\begin{deluxetable}{lccrccccccc}
\tablenum{2}
\tablewidth{0pt}
\tablecaption{Observations of Sample Galaxies \label{galxd2}}
\tablehead{
\colhead{Name} 
& \colhead{UV Date}
& \colhead{UV(Filter)\tablenotemark{a}}
& \colhead {Optical Date}
& \colhead{Telescope\tablenotemark{b}}
& \colhead{$U$\tablenotemark{a}}
& \colhead{$B$\tablenotemark{a}}
& \colhead{$V$\tablenotemark{a}}
& \colhead{$R$\tablenotemark{a}}
& \colhead{$I$\tablenotemark{a}}
}
\startdata
NGC 925 & 950314 & 1591 (B5) & 990207 & Pal1.5 & \nodata & \nodata & \nodata& 300\tablenotemark{c} & \nodata \nl
 &  &  & 990213 & Pal1.5 & 3600\tablenotemark{c} & 600 & 600 & \nodata & \nodata \nl
NGC 1068 & 950307 & 753 (B5) & 990214 & Pal1.5 & 3600 & 300 & 300 & 200 & \nodata \nl
NGC 1097\tablenotemark{d} & 950312 & 1121 (B5) & 951221 & LCO2.5 & 3x400 & 3x200 & 3x200 & \nodata & 3x200 \nl
NGC 1313\tablenotemark{d} & 950312 & 1071 (B5) & 951224 & LCO2.5 & 2x400 & 2x200 & 2x200 & \nodata & 2x200 \nl
NGC 1365\tablenotemark{d} & 950315 & 974 (B5) & 951223 & LCO2.5 & 2x400 & 2x200 & 2x200 & \nodata & 2x200 \nl
NGC 1512\tablenotemark{d}/10\tablenotemark{e} & 950315 & 949 (B5) & 951220 & LCO2.5 & 3x400 & 3x200 & 3x200 & \nodata & 3x200 \nl
NGC 1566 & 950316 & 1391 (B5) & 951222 & LCO2.5 & 2x400 & 2x200 & 2x200 & \nodata & 2x200 \nl
NGC 1672 & 950307 & 927 (B5) & 951219 & LCO2.5 & 4x400 & 3x200 & 3x200 & \nodata & 3x200 \nl
NGC 2403\tablenotemark{d} & 950308 & 772 (B1) & 971031& Pal1.5 & 1600 & 600 & 300 & \nodata & 300 \nl
 &  &  & 990208& Pal1.5 & \nodata & \nodata & \nodata & 60& \nodata \nl
NGC 2841 & 950308 & 1021 (B1) & 940214 & Pal5.0 & \nodata & 300 & \nodata & \nodata & 120,2x60 \nl
 &  & & 990213 & Pal1.5 & 3600 & \nodata & 300 & 200 & \nodata \nl
NGC 2903 & 950307 & 549 (B1) & 951123 & Pal1.5 & \nodata & 900 & 600 & \nodata & 600 \nl
 &  & & 990208 & Pal1.5 & \nodata & \nodata & \nodata & 300 & \nodata \nl
 &  & & 990214 & Pal1.5 & 3600 & \nodata & \nodata & \nodata & \nodata \nl
NGC 3115 & 950307 & 1081 (B1) & 951226 & LCO2.5 & 300 & \nodata & 300 & \nodata & 300 \nl
NGC 3226/7\tablenotemark{e} & 950316 & 1271 (B1) & 960218 & Pal1.5 & \nodata & 450 & \nodata & \nodata & \nodata \nl
 &  &  & 990207 & Pal1.5 & \nodata & \nodata & \nodata & 300 & \nodata \nl
IC 2574\tablenotemark{d} & 950316 & 624 (B1) & 960414 & Pal5.0 & \nodata & \nodata & 3x600 & \nodata & 5x400\tablenotemark{c} \nl
 &  &  & 990208 & Pal1.5 & \nodata & \nodata & \nodata & 300\tablenotemark{c} & \nodata \nl
 &  & & 990214 & Pal1.5 & 3600 & 600 & \nodata & \nodata & \nodata \nl
NGC 3310 & 950311 & 1131 (B1) & 990207 & Pal1.5 & \nodata & \nodata & \nodata & 300\tablenotemark{c} & \nodata \nl
 &  &  & 990214 & Pal1.5 & \nodata & \nodata & 300 & \nodata & \nodata \nl
NGC 3351 & 950306 & 881 (B1) & 960118 & Pal1.5 & \nodata & 600 & 300 & \nodata & 2x600,300\nl
 &  &  & 990208 & Pal1.5 & \nodata & \nodata & \nodata & 300\tablenotemark{c} & \nodata \nl
NGC 3379/84/89\tablenotemark{e} & 950306 & 1301 (B1) & 990208 & Pal1.5 & \nodata & \nodata & \nodata & 120 & \nodata \nl
NGC 4038/9\tablenotemark{d} & 950307 & 881 (B1) & 951222 & LCO2.5 & 1200 & \nodata & 600 & \nodata & 600 \nl
NGC 4151/6 \tablenotemark{e} & 950313 & 833 (B1) & 990207 & Pal1.5 & \nodata & \nodata & \nodata & 300 & \nodata \nl
NGC 4214 & 950313 & 1061 (B1) & 960514 & Pal1.5 & \nodata & 600 & 2x300 & \nodata & 300\tablenotemark{c} \nl
 &  &  & 990208 & Pal1.5 & \nodata & \nodata & \nodata & 300\tablenotemark{c} & \nodata \nl
NGC 4449 & 950307 & 987 (B1) & 960515 & Pal1.5 & \nodata & 600 & 300 & \nodata & 300 \nl
 &  &  & 990207 & Pal1.5 & \nodata & \nodata & \nodata & 300 & \nodata \nl
NGC 4647/9\tablenotemark{e} & 950311 & 1301 (B1) & 990207 & Pal1.5 & \nodata & \nodata & \nodata & 300 & \nodata \nl
NGC 4736 & 950312 & 1041 (B1) & 990208 & Pal1.5 & \nodata & 2x600 & \nodata & 300\tablenotemark{c} & \nodata \nl
 &  &  & 990213 & Pal1.5 & 3600 & \nodata & \nodata & \nodata & \nodata \nl
 &  &  & 990320 & Pal1.5 & \nodata & \nodata & 300 & \nodata & \nodata \nl
NGC 5055\tablenotemark{d} & 950315 & 1141 (B1) & 990213 & Pal1.5 & 3600 & 600 & 300 & 200 & \nodata \nl
NGC 5194\tablenotemark{d} & 950312 & 1101 (B1) & 980423 & Pal1.5 & 6x1800 & \nodata & \nodata & \nodata & \nodata \nl
 &  &  & 990207 & Pal1.5 & \nodata & 900 & 600 & 300 & 300 \nl
NGC 5236\tablenotemark{d} & 950306 & 819 (B1) & 960623 & CTIO1.5 & 4x300 & 4x300 & \nodata & 4x300 & 4x60 \nl
NGC 5253 & 950307 & 727 (B5) & 951222 & LCO2.5 & \nodata & 300 & \nodata & \nodata & \nodata \nl
 &  &  & 951223 & LCO2.5 & \nodata & \nodata & 3x600 & \nodata & 2x600  \nl
NGC 5457\tablenotemark{d} & 950311 & 1311 (B1) & 960515 & Pal1.5 & \nodata & 600 & 300 & \nodata & 300 \nl
\vspace*{-0.15in}
\tablenotetext{a}{Exposure time in seconds. (UV filter in parentheses). }
\tablenotetext{b}{LCO2.5 = Las Campanas 2.5m, CTIO1.5 = Cerro Tololo 1.5m, Pal5.0 = Palomar 5m, Pal1.5 = Palomar 1.5m.}
\tablenotetext{c}{Calibration estimated from other galaxies observed on the same night (see Section~2.4).}
\tablenotetext{d}{Optical sky background uncertain because galaxy fills the frame.}
\tablenotetext{e}{Galaxies imaged on same frame.}
\enddata
\end{deluxetable}
\end{document}